\documentclass[pre,showpacs,twocolumn,twoside]{revtex4}
\usepackage{graphicx}
\usepackage{amsmath,amssymb}

\begin{document}

\def\ket#1{|#1\rangle}
\def\bra#1{\langle#1|}
\def\scal#1#2{\langle#1|#2\rangle}
\def\matr#1#2#3{\langle#1|#2|#3\rangle}
\def\dalpha{{\dot{\alpha}}}
\def\ddalpha{{\ddot{\alpha}}}
\def\rb#1{\left(#1\right)}
\def\sb#1{\left[#1\right]}
\def\ti#1{\mathrm{#1}}													
\def\a2#1{\alpha_{#1}}
\def\aR2#1{\alpha_{#1}^{R}}
\def\aI2#1{\alpha_{#1}^{I}}
\def\p2#1{\pi_{#1}}
\def\pR2#1{\pi_{#1}^{R}}
\def\pI2#1{\pi_{#1}^{I}}
\def\da2#1{\dalpha_{#1}}
\def\dda2#1{\ddalpha_{#1}}
\def\daR2#1{\dalpha_{#1}^{R}}
\def\daI2#1{\dalpha_{#1}^{I}}
\def\Js{\mathrm{J}}
\def\poisson#1#2{\left\{#1,#2\right\}}
\def\Linvariant{\langle L\rangle}
\def\Vinvariant{\langle V\rangle}
\def\Expect#1{\langle{#1}\rangle}
\def\komut#1#2{\left[{#1},{#2}\right]}
\def\abs#1{\left\lvert#1\right\rvert}
\def\abss#1{\abs{#1}^{2}}
\newcommand{\ud}{\mathrm{d}}
\newcommand{\MeV}{\mathrm{MeV}}
\newcommand{\e}{\mathrm{e}}
\newcommand{\ui}{\mathrm{i}}
\renewcommand{\Im}{\mathop{\text{Im}}}
\renewcommand{\Re}{\mathop{\text{Re}}}
\def\tproduct#1#2#3{\left[#1 \times #2\right]^{\left(#3\right)}}
\def\tproducta#1{\left[\alpha\times\alpha\right]^{\left(#1\right)}}
\def\tproductad#1{\left[\dot{\alpha}\times\dot{\alpha}\right]^{\left(#1\right)}}
\def\abs#1{\left|#1\right|}
\newcommand{\Freg}{F_{\mathrm{reg}}}
\newcommand{\freg}{f_{\mathrm{reg}}}
\newcommand{\Jmax}{J_{\mathrm{max}}}

\title{Quantum chaos in the nuclear collective model: I. Classical-quantum correspondence}

\date{\today}
\author{Pavel Str{\' a}nsk{\' y}, Petr Hru{\v s}ka, Pavel Cejnar}
\affiliation{Institute of Particle and Nuclear Physics, Faculty of Mathematics and Physics, Charles University, 
             V~Hole{\v s}ovi{\v c}k{\' a}ch 2, 180\,00 Prague, Czech Republic}
\begin{abstract}
Spectra of the geometric collective model of atomic nuclei are analyzed to identify chaotic correlations among nonrotational states.
The model has been previously shown to exhibit a high degree of variability of regular and chaotic classical features with energy and control parameters.
Corresponding signatures are now verified also on the quantum level for different schemes of quantization and with a variable classicality constant.
\end{abstract}

\maketitle

\section{Introduction}\label{sec:Introduction}

What are typical features of a quantum system whose classical limit is chaotic?
This is a central question of so-called \lq\lq quantum chaos\rq\rq\ \cite{LH89,Gut90,Rei92,Haa92,Sto99}, a branch of quantum physics that has been attracting a considerable interest since 1970's.
Apparently, quantum systems show no trajectories, hence no Lyapunov exponents, Poincar{\'e} sections or other signatures constitutional for the distinction of chaos on the classical level.
Instead, some genuinely quantum attributes of the system seem to absorb the information on the regular or chaotic character of the classical dynamics.
The best known examples are correlation properties of the spectra of energy levels.
As a surprise, quantum systems with a chaotic classical counterpart show highly correlated quantum spectra, described within the theory of Gaussian matrix ensembles \cite{Bro81}, while the spectra of systems that are classically regular look more or less random.

In recent years, alternative signatures of quantum chaos have been proposed, like the morphology of wave functions \cite{LH89,Gut90,Sto99}, fluctuations of the scattering matrix \cite{Sto99}, sensitivity to perturbations \cite{Gor06} etc.
The research of these issues is by far not completed.
Note that the absence of an exact definition of chaos on the quantum level led to a proposal to use the term \lq\lq quantum chaology\rq\rq\ instead of quantum chaos \cite{Ber87}.

The relation of spectral properties of chaotic quantum systems to those of Gaussian matrices was proposed by Bohigas, Giannoni, and Schmit in 1984 \cite{Boh84}.
Since then, the conjecture has been tested in numerous concrete systems and supported by several involved theoretical analyses.
Recently, correlation properties of quantal spectra were rephrased into the language of stochastic time series with $\sim 1/f^{\alpha}$ type of noise, the chaotic case being identified with $\alpha=1$ \cite{fnoise}.

In spite of this progress, some problems concerning the relation of the level statistics to classical chaos remain open.
The following two questions, in particular, helped to guide the work presented in this article:
First, if the classical dynamics exhibits abrupt transitions between dominantly regular and dominantly chaotic types of motions with varying energy, to what extent does the level statistics within a single spectrum follow these changes?
Second, since the quantization is not a unique procedure, does Bohigas' conjecture hold in all quantum realizations of the given classical system?

The model we use to probe the above questions is the geometric model of nuclear collective motions \cite{Boh98}.
Classical dynamics generated by this model was recently shown to exhibit an immense variability of the dynamical modes \cite{Cej04,Str06}.
The rise of ordered modes from the chaotic ones and their breakdown are phenomena observed at numerous places in the plane of energy versus control parameter---and in a majority of cases they are not properly understood yet.
Very similar properties have been assigned also to the interacting boson model \cite{Alh91,Whe93,Mac07}, which is closely related to the geometric collective model.
A detailed analysis of dynamical features of these models is an interesting and important subject in the context of nuclear physics.
However, we consider these models to be very well suited also for more generally oriented studies, such as those seeking answers to the above questions.

The classical analysis of chaos in the geometric collective model has been presented in Refs.~\cite{Cej04,Str06}.
In the present paper, we focus on the analysis of quantum properties.
We restrict ourselves to a subset of quantum levels with zero angular momentum, which makes the configuration space effectively two-dimensional, in contrast to five-dimensional space corresponding to general motions.
It is then shown that the classical version of the model for zero rotations can be quantized in two physically meaningful ways.
We solve the eigenvalue problem in both cases and compare the level statistics obtained, looking particularly into the regions where transitions between regular and chaotic dynamics take place.
The possibility of changing the value of a classicality (Planck) constant enables us to populate the spectrum with variable density of quantum states, which is used for a global inspection of large energy domains and zooming in some finer details.

The previous paragraph outlined the content of the present part of this paper, further referred to as Part~I.
In the forthcoming part, Part~II \cite{II}, the method invented by Peres \cite{Per84} will be applied to the geometric model.
The method, which exploits specific information on the structure of individual eigenstates, makes it possible to draw the spectrum of a quantum system in a way that allows one to visually allocate regular and chaotic domains.
A close relation to classical dynamics, in particular an analogy with the graphical method based on Poincar{\'e} sections, will be demonstrated.

The plan of the present part of the paper is as follows:
In Sec.~\ref{sec:Model} (and in Appendix~\ref{sec:MatrixElements}), we introduce the model with its alternative quantizations and discuss some technical issues related to its numerical solution.
The method used to evaluate the spectral statistics is described in Sec.~\ref{sec:SpectralStatistics}.
In Sec.~\ref{sec:QuantumClassicalRelation}, the results of the statistical analysis are presented and compared with the corresponding classical measures of chaos.
Conclusions are contained in Sec.~\ref{sec:Conclusions}.

\section{The model}\label{sec:Model}

\begin{figure*}[tbp]
	\centering
	\includegraphics[width=\linewidth]{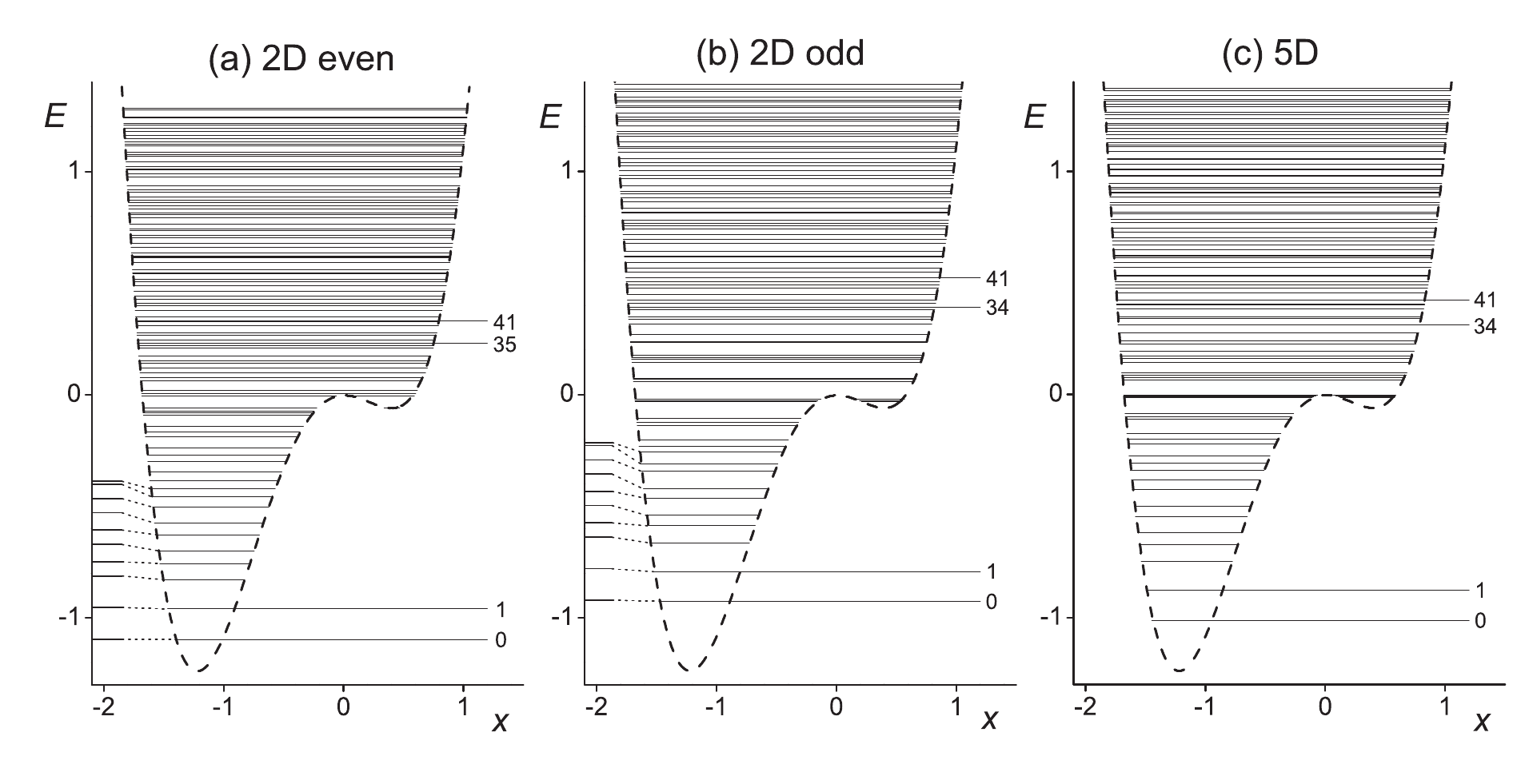}
	\caption{\protect\small 
Energy levels of the GCM Hamiltonian for $A=-1$, $B=1.09$, $C=1$, and $\hbar^{2}/K=25\cdot10^{-4}$ drawn inside the $y=0$ section of the potential well.
Energy is given in relative units.
Panels (a) and (b), respectively, show even and odd states in the 2D quantization, panel (c) corresponds to the 5D quantization.
Levels associated with the wave functions in Fig.~\ref{fig:WF} are marked with their respective ordinal numbers.
In both 2D cases, levels of the harmonic-well approximation are drawn on the left.
					}
	\label{fig:Levels}
\end{figure*}

\begin{figure*}[tbp]
\includegraphics[width=0.65\linewidth]{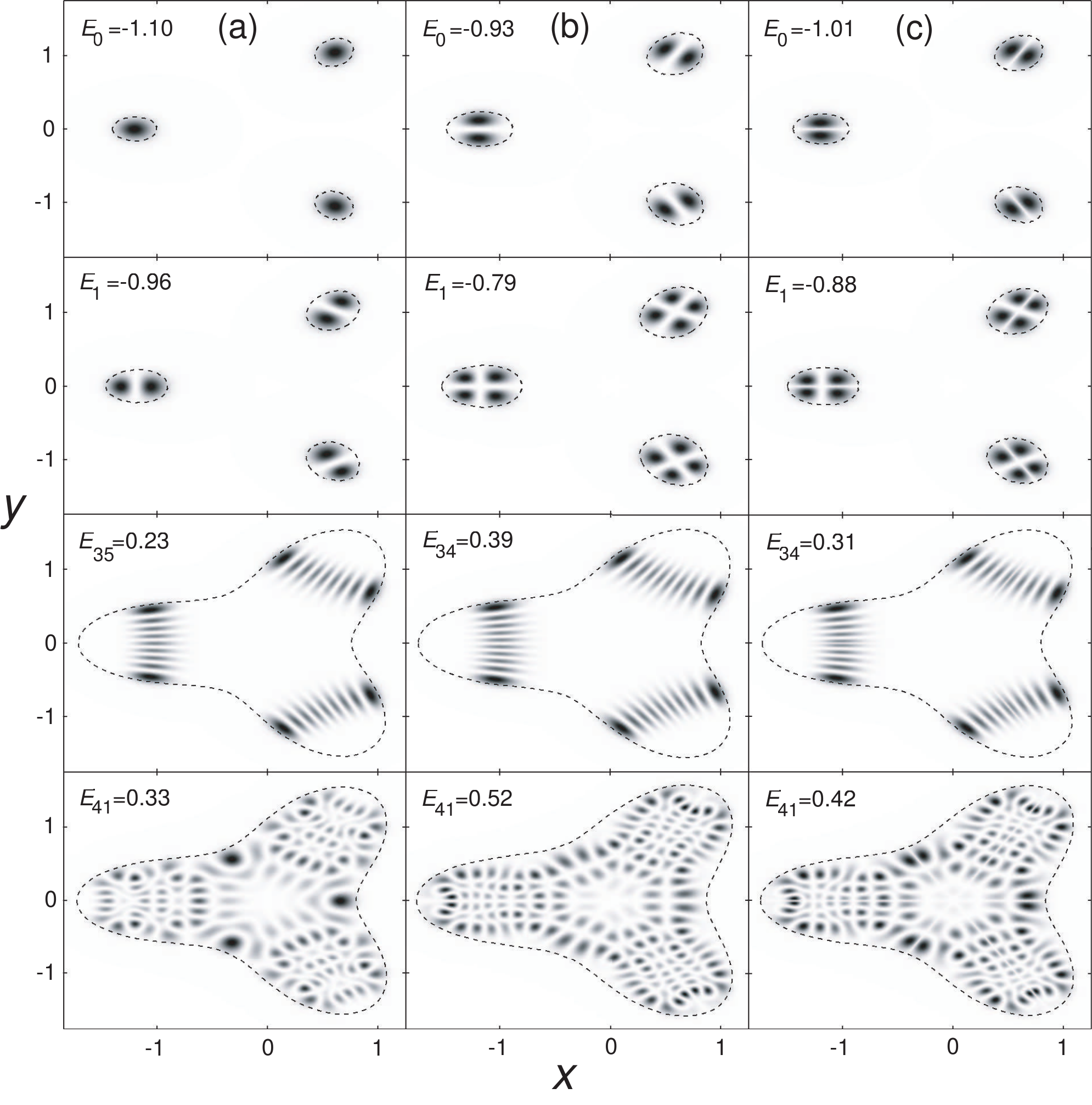}
	\caption{\protect\small 
Probability densities derived from the wave functions for selected levels from Fig.~\ref{fig:Levels}.
Columns (a), (b), and (c) show the 2D even, 2D odd, and 5D cases, respectively.
All distributions are constrained by the three-fold symmetry of the potential.
The dashed lines demarcate the kinematically accessible area at given energy; cf. the dashed lines in Fig.~\ref{fig:Levels}.
The states in two upper rows belong to the region where the quadratic-well approximation of the potential is valid.
The third and fourth row, respectively, depict examples of regular and chaotic states with higher energies.
					}
	\label{fig:WF}
\end{figure*}

\subsection{5D and 2D Hamiltonians}\label{sec:Hamiltonians}

In this section we introduce the Hamiltonian of the Geometric Collective Model (GCM) for zero rotations, $J=0$, and provide two different ways of its quantization, which are referred to as five-dimensional (5D) and two-dimensional (2D) cases.
		
The kinetic and potential terms of the GCM Hamiltonian $H=T+V$,
\begin{eqnarray}
T&=&\frac{\sqrt{5}}{2K}\tproduct{\pi}{\pi}{0}
\,,\label{eq:Tscal}\\
V&=&\sqrt{5}A\tproduct{\alpha}{\alpha}{0}-\sqrt{\tfrac{35}{2}}B\tproduct{\tproduct{\alpha}{\alpha}{2}}{\alpha}{0}\nonumber
\nonumber\\
&&+5C\left(\tproduct{\alpha}{\alpha}{0}\right)^{2}
\,,\label{eq:Vscal}
\end{eqnarray}
are built from generalized complex coordinates $\alpha\equiv\alpha^{(2)}_{\mu}$ (with $\mu=-2,\dots,+2$) and the corresponding conjugate momenta $\pi\equiv\pi^{(2)}_{\mu}$, which are both spherical tensors of rank 2.
Parameter $K$ in the kinetic part has the meaning of mass, while $\{A,B,C\}$ determine the form of the potential.
Note that $\tproduct{a}{b}{\lambda}$ stands for a coupling of general tensors $a$ and $b$ to angular momentum $\lambda$.
The Hamiltonian is rotationally invariant since it contains only scalar couplings of coordinates and momenta. 
		
The model is usually approached via an expansion of the nuclear radius into spherical harmonics, $R=R_0(1+\sum_{\lambda\mu}\alpha^{(\lambda)*}_{\mu}Y^{(\lambda)}_{\mu})$, with only the $\lambda=2$ terms taken into account; hence the name geometric model.
However, the coordinates can also have different interpretations keeping only their quadrupole tensor character.
The geometric Hamiltonian in the above form was introduced by Aage Bohr in 1952 \cite{Boh52}.
A way of systematic construction of higher-order terms in both potential and kinetic parts of the Hamiltonian was presented by Gneuss, Mosel, and Greiner \cite{GCM}.
Several other types of potential have been considered in connection with shape transitions in nuclei.
An overview of these potentials with relevant references and the corresponding quantum solutions can be found in Ref.~\cite{For05}.

Coordinates $\alpha$ satisfy the constraint $\alpha^{(2)*}_{\mu}=(-)^{\mu}\alpha^{(2)}_{-\mu}$ and therefore contain five independent real variables.
Two of these variables capture the intrinsic shape of the nucleus (with a quadrupole deformation) and the remaining three variables describe the orientation of the nucleus in the laboratory frame (they can be associated with the Euler angles transforming the lab frame to the intrinsic one).
In the intrinsic frame, only the shape variables are relevant.
These are connected with the two independent scalar combinations of $\alpha$'s in Eq.~(\ref{eq:Vscal}) and are usually parametrized as follows: 
\begin{equation}
\alpha^{(2)}_{0}\equiv\beta\cos\gamma\equiv x\,,\quad
\sqrt{2}\Re\alpha^{(2)}_{\pm 2}\equiv\beta\sin\gamma\equiv y
\end{equation}
($\alpha^{(2)}_{\pm 1}=0$).
This yields the potential in the form:
\begin{equation}\label{eq:V}
V=A\beta^{2}+B\beta^{3}\cos{3\gamma}+C\beta^{4}
\,.
\end{equation}

Standard quantization procedure with $\pi_{\mu}=-i\hbar\tfrac{\partial}{\partial\alpha_{\mu}}$ 
leads to the kinetic term \cite{Boh52}
\begin{eqnarray}
T^{\rm 5D}=&-&\frac{\hbar^{2}}{2K}\left(\frac{1}{\beta^{4}}\frac{\partial}{\partial\beta}\beta^{4}\frac{\partial}{\partial\beta}
+\frac{1}{\beta^{2}\sin{3\gamma}}\frac{\partial}{\partial\gamma}\sin{3\gamma}\frac{\partial}{\partial\gamma}\right)
\nonumber\\
&+&T_{\rm rot}
\,,
\label{eq:T5D}
\end{eqnarray}
where $T_{\rm rot}$ stands for a nontrivial rotational part of the kinetic energy, containing the derivatives with respect to Euler angles and 
coupling all five dynamical variables.
Since we restrict ourselves only to nonrotating regimes, $J=0$, we set $T_{\rm rot}=0$.
This can be seen as a projection of the full 5D coordinate system into an effectively two-dimensional space describing vibrational degrees of freedom $\beta$ and $\gamma$.
Note, however, that this projection differs from the 2D case by a modified definition of the scalar product, namely:
\begin{equation}\label{eq:scal}
\scal{\Psi_1^{\rm 5D}}{\Psi_2^{\rm 5D}}=\int_{0}^{2\pi}\int_{0}^{\infty}\Psi_1^{{\rm 5D}*}\Psi_2^{\rm 5D}\underbrace{\beta^{4}\abs{\sin{3\gamma}}}_{{\cal N}}d\beta\,d\gamma
\,,
\end{equation}
where $\Psi_{i}^{\rm 5D}(\beta,\gamma)$, with $i=1,2$, stands for two wave functions and ${\cal N}(\beta,\gamma)$ represents a measure. 
Appendix~\ref{sec:MatrixElements} gives some details, see in particular Eq.~\eqref{eq:5DNormalization}.
The normalization condition $\scal{\Psi^{\rm 5D}}{\Psi^{\rm 5D}}=1$ leads to another class of solutions than would apply in case of a genuinely 2D Hamiltonian.

For $J=0$, however, the truly 2D scheme represents an alternative way of quantization.
It means that we put the system into the intrinsic frame first and only then carry out the quantization.
The kinetic term obtained in this way reads as follows:
\begin{equation}\label{eq:T2D}
T^{\rm 2D}=-\frac{\hbar^{2}}{2K}\left(\frac{1}{\beta}\frac{\partial}{\partial\beta}\beta\frac{\partial}{\partial\beta}
+\frac{1}{\beta^{2}}\frac{\partial^{2}}{\partial\gamma^{2}}\right)
\,,
\end{equation}
which is nothing but the standard 2D kinetic energy expressed in polar coordinates $(r,\phi)\equiv(\beta,\gamma)$.
The scalar product in this case is defined in the usual way, therefore the space of solutions coincides with $L^2(\mathbb{R}^2)$.
Note that the Hamiltonian given by Eqs.~\eqref{eq:T2D} and \eqref{eq:V} is a generalization of the widely studied H{\' e}non-Heiles	model \cite{Hen64}.
Let us stress that in the nuclear physics context only the 5D quantization is correct.

Both forms \eqref{eq:T5D} and \eqref{eq:T2D} have the same classical limit for $J=0$.
The corresponding Hamiltonians $H^{\rm 5D}$ and $H^{\rm 2D}$, respectively, with the common potential \eqref{eq:V}, enable one to study the impact of the quantization method on spectral correlations.
Although individual energy eigenvalues obtained in both quantizations are different, we may assume---as implicit in Bohigas' conjecture---that the spectral statistics remains essentially the same (after correctly separating levels with different conserved quantum numbers in both cases).
The validity of this assumption will be discussed in Sec.~\ref{sec:QuantumClassicalRelation}.
Note that despite the 2D and 5D quantum Hamiltonians carry a clear physical meaning, they represent just two options out of an infinite number of quantization possibilities.

\subsection{Numerical solution}\label{sec:Diagonalization}

Both versions of the $J=0$ GCM Hamiltonian are diagonalized numerically, using the eigenbases of a 5D or 2D harmonic oscillator.
The oscillator Hamiltonian reads as $H^{\bullet}_{\rm osc}=T^{\bullet}+V_{\rm osc}$, where $V_{\rm osc}=A_{\rm osc}\beta^{2}$ (with $A_{\rm osc}$ being an arbitrary positive constant whose optimal choice will be discussed later) and $T^{\bullet}$ stands for the 5D or 2D kinetic operator \eqref{eq:T5D} or \eqref{eq:T2D}, respectively (we denote $\bullet=$5D or 2D).
A general GCM Hamiltonian $H^{\bullet}=T^{\bullet}+V$ is expressed as $H^{\bullet}=H^{\bullet}_{\rm osc}+V'$, where $V'=V-V_{\rm osc}$.
It turns out that matrix elements of $V'$ in both oscillator bases can be expressed analytically, which makes the process of numerical diagonalization very efficient.
Details and explicit expressions can be found in Appendix~\ref{sec:MatrixElements}.

The original 5D solution of the GCM possesses several implicit symmetries, namely
\begin{align}\label{eq:WFSymmetry1}
\Psi^{\ti{5D}}(\beta,\gamma)&=\Psi^{\ti{5D}}\left(\beta,\gamma+\tfrac{2\pi}{3}\right)
\,,\\
\label{eq:WFSymmetry2}
\Psi^{\ti{5D}}(\beta,\gamma)&=\Psi^{\ti{5D}}(\beta,-\gamma)
\,,
\end{align}
with $\Psi^{\ti{5D}}$ an arbitrary wave function in 5D.
These relations arise from the ambiguity of the system's orientation in the intrinsic frame.
On the other hand, solutions of the 2D model do not a priori satisfy such symmetries.
If the spectra associated with both quantizations are to be compared, conditions \eqref{eq:WFSymmetry1} and \eqref{eq:WFSymmetry2} need to be imposed externally also to the 2D case.
This is done by selecting a subset $\Psi^{\ti{2D}}_{\ti{E},nm}$ (where E stands for even in variable $\gamma$) of the 2D oscillator basis in which the diagonalization is carried out (see Appendix~\ref{sec:MatrixElements}).
In addition, if we relax condition \eqref{eq:WFSymmetry2} and require only the equality of absolute values of the wave functions involved, 
we can take into account another independent class of 2D solutions, namely the wave functions odd in variable $\gamma$, i.e. satisfying $\Psi^{\ti{2D}}(\beta,\gamma)=-\Psi^{\ti{2D}}(\beta,-\gamma)$.
These are obtained by diagonalization in the subset $\Psi^{\ti{2D}}_{\ti{O},nm}$ of the 2D oscillator states (Appendix~\ref{sec:MatrixElements}).

The Hamiltonian matrix expressed in the truncated oscillator basis has a band form.
The band width is approximately equal to the maximal value of the principal quantum number in the selected subset of basis states.
This makes the diagonalization feasible even at relatively high dimensions.
The convergence of solutions is checked by a visual inspection of the distribution of eigenvector components in the oscillator basis (a bad convergence is signaled by missing tails of the computed distributions) and/or by trial calculations using variable size of the basis (we test the stability of computed eigenvalues against an increase of the dimension).
Our procedure guarantees that the precision $\delta E$ of individual eigenvalues satisfies the condition $\delta E\ll\Delta E$, where $\Delta E$ is an average spacing between levels in the selected part of the spectrum.
This is needed for the determination of the nearest-neighbor spacing distribution.

In order to make the net spectra of converged eigensolutions as large as possible, we optimize the oscillator parameter $A_{\rm osc}$ that determines a characteristic scale of the basis wave functions.
The procedure is based on the determination of the $A_{\rm osc}$ value for which the trace of the GCM Hamiltonian in the truncated basis is minimal (the optimal choice of $A_{\rm osc}$ is however lower than this value, as empirically verified for the parameter ranges studied here).
Taking all these issues into account, we have found that on a common personal computer one can employ up to $10^5$ basis states and obtain up to about $5\cdot 10^4$ well converging eigensolutions (exact numbers still depend on the choice of external parameters).

An example showing all three classes of solutions---i.e., 2D even, 2D odd, and 5D---is plotted in Fig.~\ref{fig:Levels}.
Here, parameters of Eq.~\eqref{eq:V} were chosen such that the potential has a minimum at $\beta\neq 0$.
Although the spectra in the three panels of Fig.~\ref{fig:Levels} look different, very close similarities become apparent when comparing also the corresponding wave functions.
This is done for selected levels in Fig.~\ref{fig:WF}.
At very low energies, the system is fully regular since the minimum of the potential \eqref{eq:V} can be approximated by a quadratic well.
Wave functions belonging to this region are seen in the	first and second rows of Fig.~\ref{fig:WF}.
On the other hand, the region of mixed dynamics is exemplified by wave functions in the third and fourth rows, which correspond to regular and chaotic cases, respectively.
Indeed whereas wave functions in the third row exhibit regular behavior (the wave function is localized within an area following some specific classical periodic orbits \cite{Sto99}), wave functions in the fourth row show diverse structures and cover the whole accessible area.

It should be noted that the 2D and 5D cases differ in the differential element needed to calculate the probability distribution in the $x\times y$ plane.
In the 2D quantization we simply have $|\Psi^{\rm 2D}(x,y)|^2dx\,dy=|\Psi^{\rm 2D}(\beta,\gamma)|^2|{\cal J}|d\beta\,d\gamma$, where ${\cal J}=\beta$ is the Jacobian of the transformation from $(x,y)$ to $(\beta,\gamma)$.
The first two columns of Fig.~\ref{fig:WF} show just the squared modulus of the respective wave functions. 
In the 5D case, however, all matrix elements contain the measure ${\cal N}$ from Eq.~\eqref{eq:scal}.
In the rightmost column of Fig.~\ref{fig:WF} we show the squared wave function $|\Psi^{\rm 5D}(\beta,\gamma)|^2$ multiplied by a factor ${\cal N}/{\cal J}=\beta^{3}\abs{\sin{3\gamma}}$.
The resulting 5D density therefore vanishes where $\sin{3\gamma}=0$, so it is visually similar to the 2D {\em odd\/} case, in spite of the condition \eqref{eq:WFSymmetry2}.
This is also why the ground-state density in Fig.~\ref{fig:WF}(c) has two maxima in each potential well, although the wave function itself has no node.

Let us finally briefly remark on the choice of parameters in this work.
It is closely connected with the scaling properties of the GCM Hamiltonian discussed in Refs.\cite{Str06,Cap03}.
In the classical case \cite{Str06}, only one of the parameters $\{A,B,C\}$ determines the scale-independent behavior of the system, while the others and $K$ can be set to $+1$ or $\pm 1$ for $A$.
In the quantum case \cite{Cap03}, the classicality parameter $\kappa=\hbar^{2}/K$ (whose changes can be viewed either as changes of the Planck constant, or as changes of the mass) constitutes the second independent parameter of the model which cannot, in general, be scaled to unity.
This parameter determines the absolute density of states.

In the following, we take $B$ as the principal control parameter and choose $\kappa$ to locate a sufficient number of levels into the energy region of interest.
The remaining parameters are fixed to $(A,C)=(-1,+1)$, which in the nuclear context corresponds to nuclei with stable ground-state deformations.
For $B=0$, the system is completely integrable (since in this case the Hamiltonian does not depend on $\gamma$ and the $x\times y$ angular momentum is an obvious integral of motions). 
Therefore, the value of parameter $B$ represents the strength of a nonintegrable perturbation.

Let us note that all the above-introduced quantities and parameters are considered here dimensionless.
The conversion to a concrete scale requires to choose the same unit for energy $E$ and parameters $A$, $B$, $C$, and $\kappa$.
In nuclear context, the unit of $\{A,B,C\}$ is set by the form of the potential (e.g. the depth of the minimum).
The appropriate value of $\kappa$ (connected with the effective mass parameter $K$) can then be determined by adjusting the number of states in a certain interval (e.g. below $E=0$).

\section{Spectral statistics}\label{sec:SpectralStatistics}

According to Bohigas' conjecture \cite{Boh84}, chaotic systems exhibit strong correlations between levels, which result in an effect known as \lq\lq spectral rigidity\rq\rq.
The short-range component of these correlations is most clearly manifested in the distribution of the nearest-neighbor spacings (NNS), i.e. gaps between neighboring levels in a transformed (so-called unfolded) spectrum.
In fully chaotic systems, this distribution is amazingly well approximated by the Wigner distribution, while in systems with regular classical counterparts the NNS distribution tends to be Poissonian.

A suitable quantity allowing one to interpolate between the two limiting cases is the Brody parameter $\omega$ \cite{Bro73}.
It is defined through the distribution 
\begin{eqnarray}\label{eq:bro}
P(s;\omega)=(\omega+1)\alpha_{\omega}s^{\omega}\exp\left(-\alpha_{\omega}s^{\omega+1}\right)
\,,\\
\alpha_{\omega}=\left[\Gamma\left(\frac{\omega+2}{\omega+1}\right)\right]^{\omega+1}
,
\nonumber
\end{eqnarray}
where $s$ is the spacing between adjacent levels in the unfolded spectrum and $\alpha_{\omega}$ a factor resulting from the required conditions
$\int_{0}^{\infty}P(s;\omega)ds=1$ (normalization) and $\quad\int_{0}^{\infty}sP(s;\omega)ds=1$ (unfolding).
Eq.~\eqref{eq:bro} interpolates between the Poisson ($\omega=0$) and Wigner ($\omega=1$) distributions, hence a value $\omega\in[0,1]$ obtained from a concrete spectrum tells us where between order and chaos the actual system is.
In spite of an artificial character of this interpolation, it has been argued that the Brody distribution is capable of fitting the data generated by realistic systems with mixed dynamics	\cite {Che90}.

\begin{figure}[tbp]
	\centering
\includegraphics[width=\linewidth]{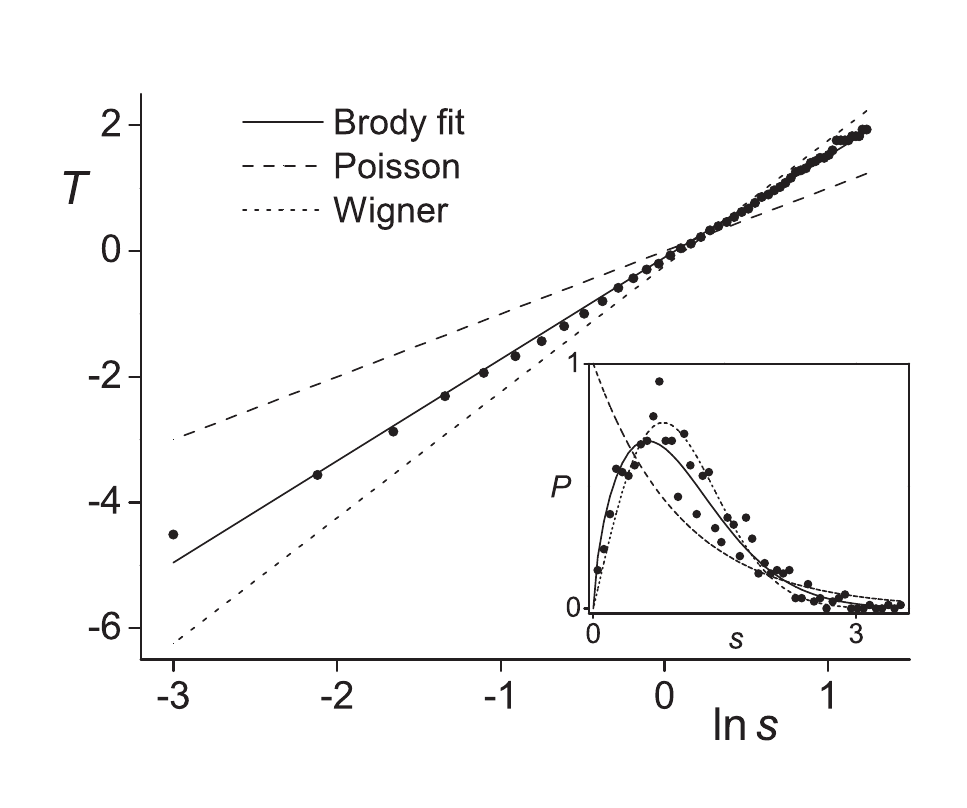}
	\caption{\protect\small 
An example of the linear fit illustrating the determination of the Brody parameter from an unfolded spectrum of levels within a single bin [it belongs to the dotted line in Fig.~\ref{fig:freg062}(b) at $E\sim1.33$].
The left-hand side of Eq.~\eqref{eq:lifit} is denoted as $T$.
The NNS distribution and its fitted Brody approximation (with $\omega=0.62$) are shown (by the full line) in the inset. 
Dashed and dotted lines correspond to Poisson and Wigner distributions, respectively.
		}
	\label{fig:lin}
\end{figure}

We use the following procedure:
Eigenstates obtained by the Hamiltonian diagonalization in a sufficiently large basis are split into groups (bins), each of them counting 1000 consecutive levels.
The standard polynomial unfolding procedure \cite{Ple02} is applied in each group, removing locally a smooth part of the level density and normalizing the average spacing to unity.
To obtain the Brody parameter, we use the identity following from Eq.~\eqref{eq:bro} \cite{Pro93}:
\begin{equation}\label{eq:lifit}
\ln\ln[1-I(s;\omega)]^{-1}=\ln\alpha_{\omega}+(1+\omega)\ln{s}
\,.
\end{equation}
Here, $I(s;\omega)=\int_{0}^{s}P(s';\omega)ds'$ can be estimated from a cumulative histogram of spacings in the unfolded spectrum.
A simple linear fit of the expression on the left-hand side of Eq.~\eqref{eq:lifit} in the logarithmic scale of variable $s$ yields the Brody parameter $\omega$ for each bin of levels.
An example is shown in Fig.~\ref{fig:lin}.
The bins subject to the above evaluations increase in energy and overlap with each other (the shift between successive bins was set to 100 levels).
The average energy of each bin is drawn on the abscissa in the resulting dependence of $\omega$ on $E$ (cf. Figs.~\ref{fig:freg109}--\ref{fig:freg062} below).

Although the linear fit \eqref{eq:lifit} can be easily implemented and demonstrates the validity of the Brody approximation over a broad domain of spacings, it may increase the relative weight of large values of $s$ in the calculation of $\omega$ \cite{Pro93}.
We have performed a numerical simulation showing that the value of the Brody parameter determined in this way may be decreased by an absolute value $\Delta\omega_{\rm syst}\approx -0.08$, while the statistical error resulting from finiteness of the sample of levels is estimated as $\Delta\omega_{\rm stat}\approx\pm 0.07$.
These uncertainties should be taken into account when evaluating the dependence of $\omega$ on energy, see Sec.~\ref{sec:QuantumClassicalRelation}.

We tried to implement also the new method \cite{fnoise} based on the $1/f^{\alpha}$ noise in spectral fluctuations, with $\alpha\in [1,2]$ corresponding to spectra in between fully chaotic ($\alpha=1$) and fully regular ($\alpha=2$) limits.
The advantage of this method lies in its simple and elegant formulation (with no explicit reference to random matrix ensembles) and in the fact that it simultaneously captures both short- and long-range spectral correlations.

In our case, however, the results were not quite satisfactory. 
The reason---an insufficient statistics---may be anticipated to be present in a majority of systems in which the competition between regular and chaotic motions quickly varies with energy.
Indeed, the work \cite{fnoise} demonstrates the power of the $1/f^{\alpha}$ method on the Robnik billiard that (as all billiard or cavity systems) exhibits a constant, energy-independent ratio between regular and chaotic phase-space volume.
This allowed the authors to average over a huge number of successive sets of levels and to get very precise results.
In contrast, properties of individual GCM trajectories cannot be trivially scaled with energy.
This feature, which in fact represents an important motivation for the detailed analysis of the present system, results in a significant increase of the statistical error of the deduced (energy-dependent) exponents $\alpha$, in some cases even exceeding 30\%.

\section{Results}\label{sec:QuantumClassicalRelation}

In this section, we will compare the quantum measure of regularity (Brody parameter) obtained in the way described in Sec.~\ref{sec:SpectralStatistics} with the corresponding classical measure.
We will focus on the influence of the different quantization schemes and on the dependence of results on the classicality parameter $\kappa$.

In our previous work \cite{Cej04,Str06}, classical measures of chaos in the geometric model were studied.
We solved the classical equations of motions for a number $N_{\rm tot}$ of trajectories with fixed energy $E$ and then, using the method based on so-called alignment indices \cite{SALI}, classified each trajectory as either regular or chaotic \cite{Str06}.
In order to quantify the overall degree of regularity at given energy, we calculated a regular fraction of the phase space $\freg=N_{\rm reg}/N_{\rm tot}$, where $N_{\rm reg}$ represents the number of regular trajectories in the sample. 
The regular fraction $\freg$ takes values from 0 (fully chaotic dynamics) to 1 (fully regular dynamics) and it can be compared with an adjunct $(1-\omega)$ of the Brody parameter.

\begin{figure}[tbp]
	\centering
	\includegraphics[width=\linewidth]{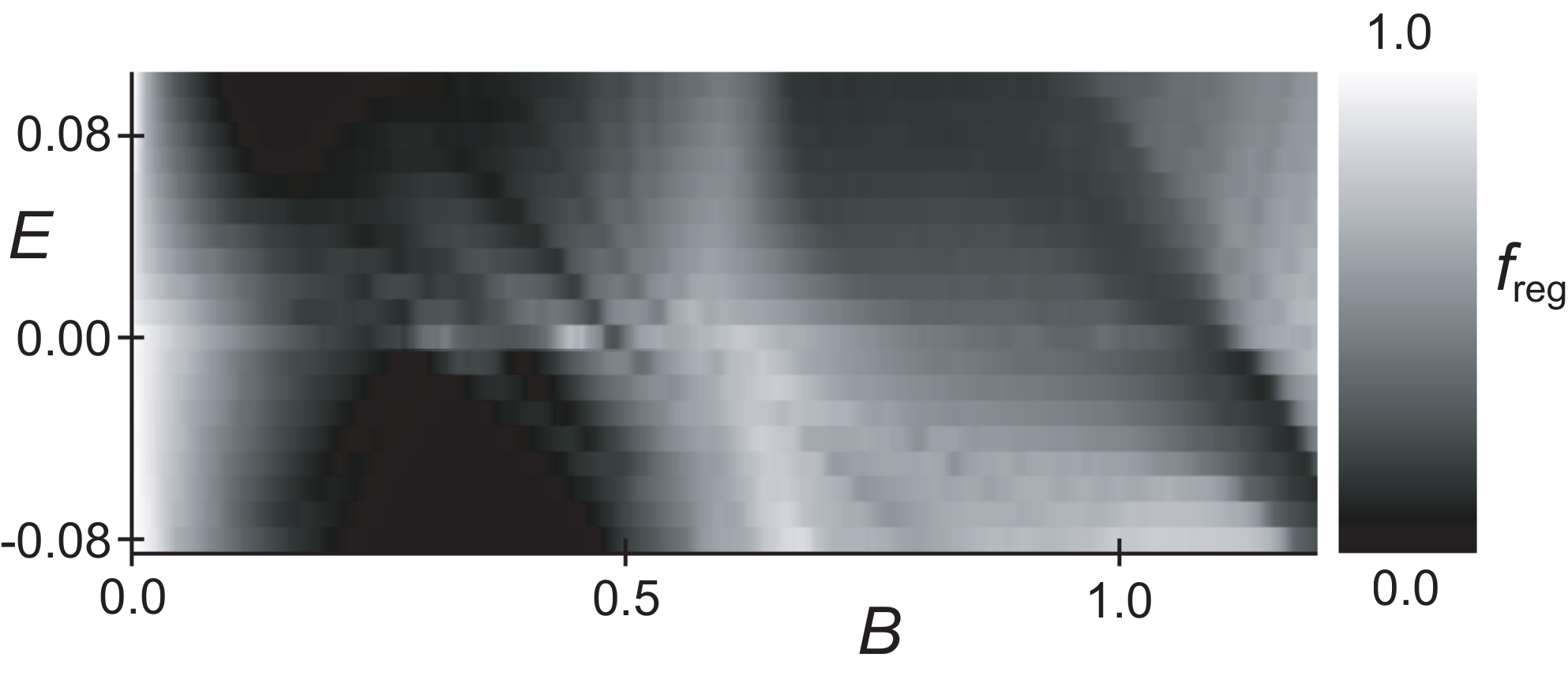}
	\caption{\protect\small 
Regular fraction $\freg$ of the classical phase space as a function of the control parameter $B$ and energy $E$ in a vicinity of $E\approx 0$.
The degree of chaos is coded in shades of gray, with white (black) corresponding to complete order (chaos).
One may notice an island with increased regularity near $B\approx 0.6$.
The number of bins in the $B$ direction is much larger than that in the $E$ direction.
		}
		\label{fig:fregDensity}
\end{figure}

In the classical case, the dependence of $\freg$ on energy and control parameter $B$ exhibits very complex nonmonotonous behavior, which is for energies around $E=0$ depicted in Fig.~\ref{fig:fregDensity}.
The following features of this dependence are worth mentioning:
First, the system is regular for small values of $B$, since $B=0$ represents a fully integrable limit of the model (see the end of Sec.~\ref{sec:Model}).
Second, a well-pronounced island of regularity in a wide range of energies is observed at $B\approx0.6$.
As shown in Ref.~\cite{Mac07}, this region is connected with the so-called regular arc of the interacting boson model \cite{Alh91,Whe93}. 
Third, increased values of the regular fraction are observed for some values of $B$ at $E\approx 0$.
The absolute energy $E=0$ corresponds to a local maximum of the potential \eqref{eq:V} at $\beta=0$.

\begin{figure}[tbp]
	\centering
	\includegraphics[width=\linewidth]{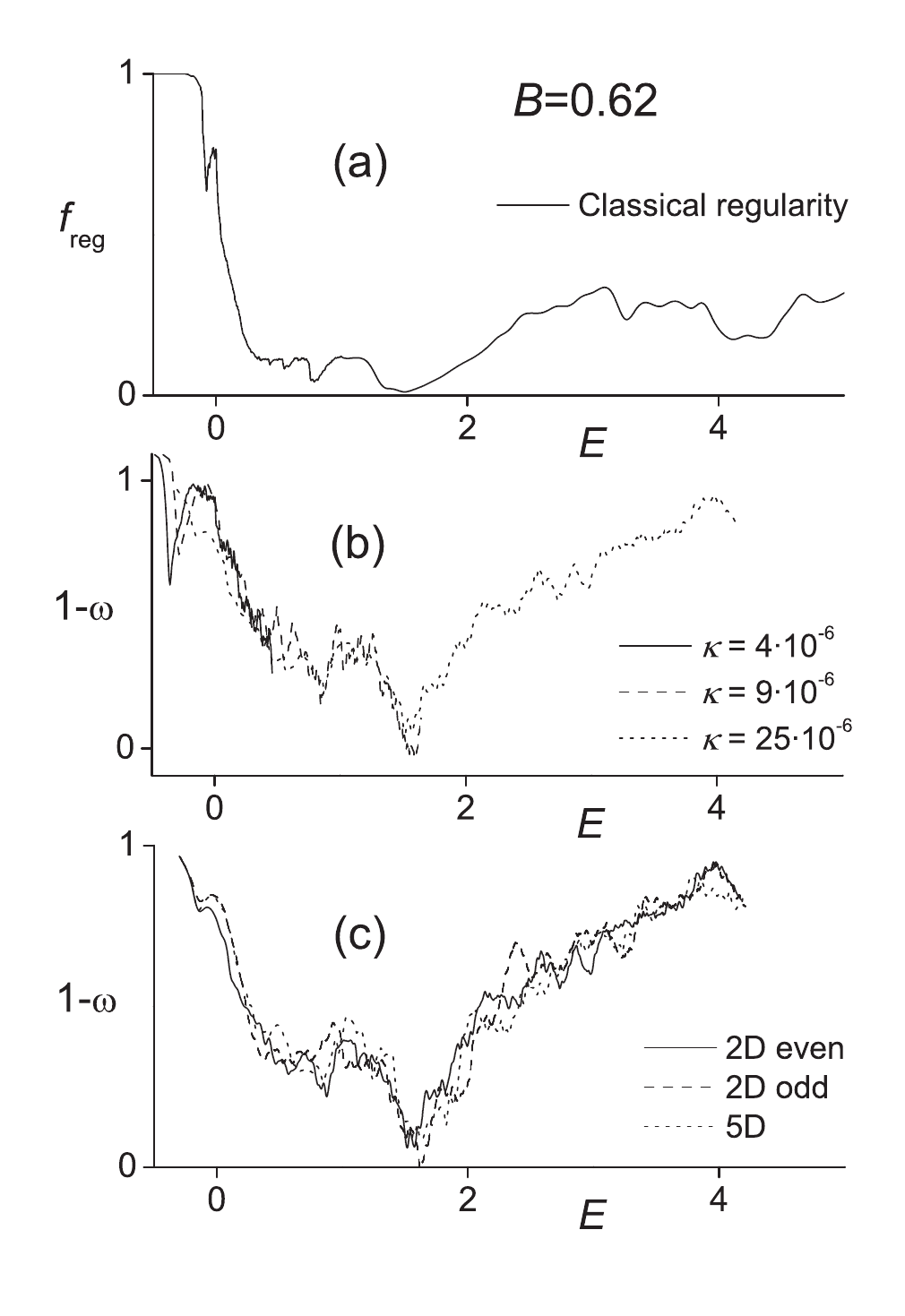}
	\caption{\protect\small 
A comparison between the classical regular fraction $\freg$ (panel a) and an adjunct $(1-\omega)$ of the Brody parameter (panels b and c) for $B=0.62$ (the main maximum of regularity at $E=0$).
Results for different values of $\kappa$ in the 2D even case are shown in panel (b), while results for different types of quantization for $\kappa=25\cdot10^{-6}$ are presented in panel (c).
A reasonable agreement of classical and quantum measures is observed.
The jittering of individual curves is caused by statistical errors.
		}
		\label{fig:freg062}
\end{figure}

\begin{figure}[tbp]
	\centering
	\includegraphics[width=\linewidth]{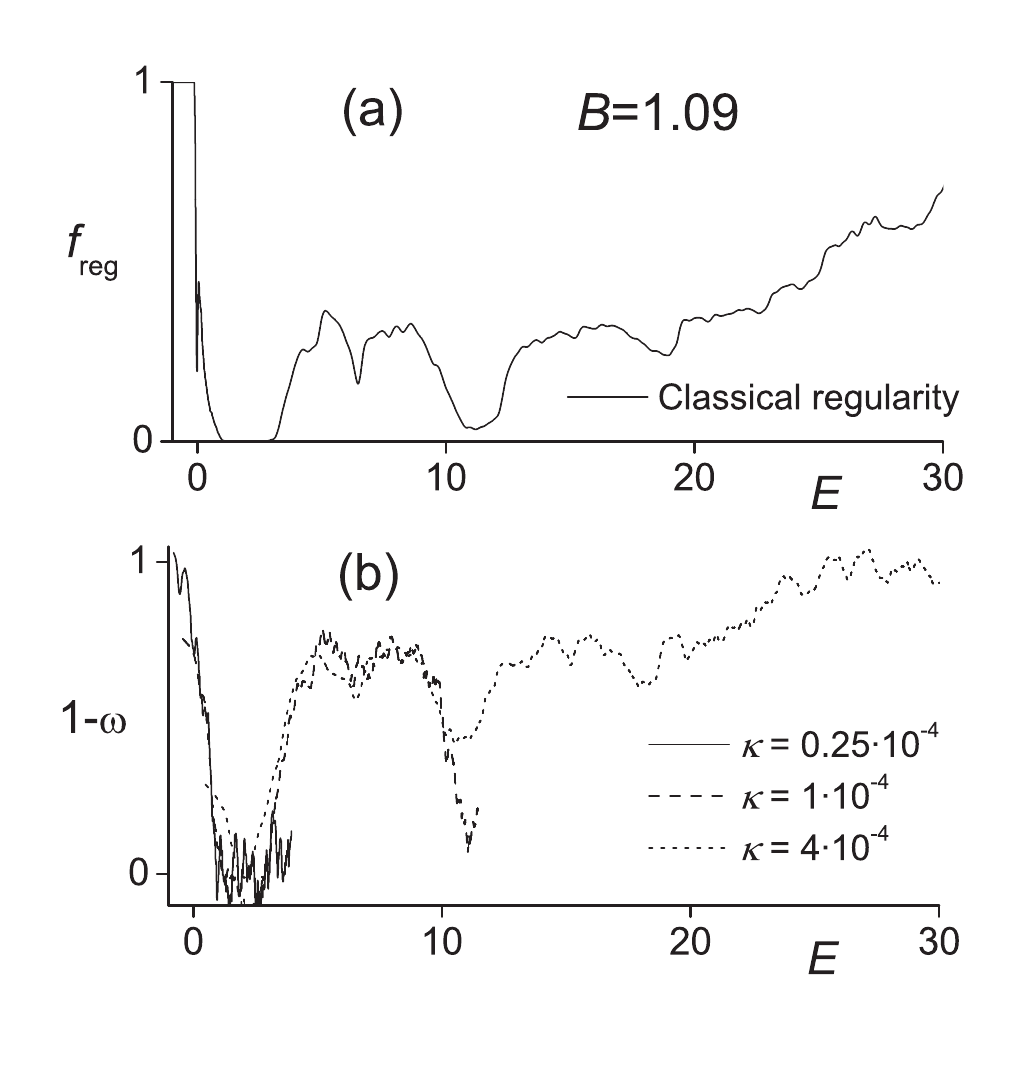}
	\caption{\protect\small 
The same as in Fig.~\ref{fig:freg062}, but for $B=1.09$ (a minimum of regularity for $E=0$).
Only the 2D even quantization is analyzed for different values of $\kappa$.
		}
		\label{fig:freg109}
\end{figure}

\begin{figure}[tbp]
	\centering
	\includegraphics[width=\linewidth]{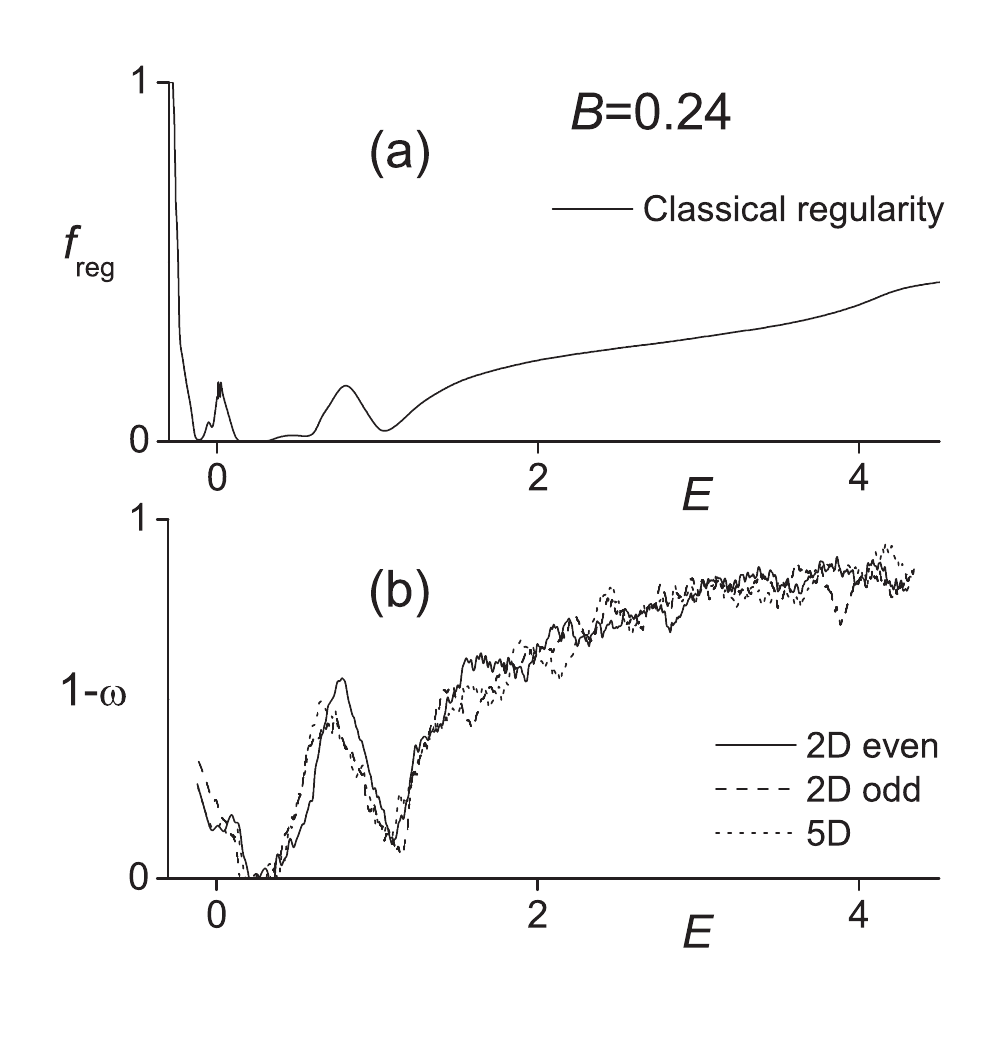}
	\caption{\protect\small 
Classical (a) and quantum (b) chaotic measures for $B=0.24$ (a minimum of regularity for $E=0$).
Only results with $\kappa=25\cdot10^{-6}$ are shown in panel (b) for various quantizations
(fluctuations are caused by statistical errors).
					}
	\label{fig:freg024}
\end{figure}

Figures~\ref{fig:freg062}--\ref{fig:freg024} depict the dependence of both measures $\freg$ and $(1-\omega)$ on energy.
The classical measure is always shown in the uppermost panel.
The value of the Hamiltonian parameter $B$ in the three figures was chosen as $B=0.62$, $B=1.09$, and $B=0.24$, respectively, with regard to the location of some extremes of regularity at $E=0$ \cite{Cej04}.
The energy range shown in the figures represents the most interesting region, lying between the domains of full regularity at very low and very high energies.
For energies just above the global minimum of the potential the system is entirely regular due to the validity of the harmonic-well approximation.	
With increasing energy, the regularity suddenly breaks down and continues falling sharply until it nearly reaches zero.
After this stage, it takes off again, somewhat surprisingly, and exhibits several well pronounced peaks of highly regular motions, which are separated by valleys of more chaotic dynamics.
These structures can be seen in Figs.~\ref{fig:freg062}--\ref{fig:freg024}.
For sufficiently high energies, not shown in the present figures, the regularity starts growing steadily toward the fully regular limit, following roughly a logarithmic dependence.
This is connected with the dominance of the $\beta^4$ term of the potential at high energies \cite{Cej04,Str06}.

The lower panels of Figs.~\ref{fig:freg062}--\ref{fig:freg024} show the corresponding quantum measure, the adjunct of the Brody parameter.
In accord with the discussion in Sec.~\ref{sec:SpectralStatistics}, we estimate the absolute errors of the $(1-\omega)$ curves as 
$-0.07$ and $+0.11$.
The reason for different error sizes in up and down directions is the above-discussed systematic error, which tends to underestimate the value of $\omega$.
On the other hand, a numerical error of $\freg$ was estimated as $\lesssim 5\%$ \cite{Str06}.
If the errors are taken into account, the fluctuations observed in the lower panels of all three figures are smoothened out and the overall correspondence between $\freg$ and $(1-\omega)$ becomes rather good.
In particular, the observed maxima and minima of both curves coincide.
It turns out that the Brody parameter tends to slightly overestimate the regularity---this being so even if the above systematic error is considered.
Note that a similar	behavior was observed for the exponent $\alpha$ in the $1/f^{\alpha}$ noise analysis \cite{fnoise}.
Indeed, there is no reason to expect that $\freg$ and $(1-\omega)$ behave in exactly the same way.
We only expect a qualitative agreement, and that is fully confirmed in the present calculation.

A remark is needed concerning the minimum seen in Fig.~\ref{fig:freg062}(b) just below $E=0$ for lower values of $\kappa$. 
It has no apparent counterpart in panel (a).
Indeed, this	minimum is only an artifact connected with the resonance of $\beta$ and $\gamma$ vibrational energies at $B=0.62$ \cite{Mac07}.
The appearance of nearly equidistant bunches of levels in the resonance region causes a serious deviation of the NNS distribution from the Brody form, which results in a nonrealistic value of the Brody parameter.
The discrepancy is localized only in a relatively narrow interval and gets washed out as $\kappa$ increases (the incriminated energy region is populated by a decreasing number of levels; a similar effect was discussed in Ref.~\cite{Hal84}).

The dependence of the quantum results on the classicality constant $\kappa$ is shown in Figs.~\ref{fig:freg062}(b) and \ref{fig:freg109}(b).
We observe that the value of $\kappa$ does not affect the energy dependence of the Brody parameter.
Mutual deviations of the curves for various $\kappa$ are bound inside the standard error interval.
Instead, the curves for distinct $\kappa$ differ in the width of the displayed energy range.
It has the following reason:
Since a decreasing value of the classicality parameter raises the density of the spectrum, the plots for smaller $\kappa$ are more detailed but cannot reach higher energies because of computational limitations of the diagonalization procedure.
In our case, $3\cdot 10^4$ reliable energy levels were calculated for each value of $\kappa$ (the dimension of the diagonalized matrix being about two times larger) and the curves are cut at the centroid energy of the uppermost bin of levels (see Sec.~\ref{sec:SpectralStatistics}).

The dependence of the quantum measure of chaos on the method of quantization is shown in Figs.~\ref{fig:freg062}(c) and \ref{fig:freg024}(b).
As was demonstrated in Fig.~\ref{fig:Levels}, the spectra obtained by the three types of quantizations (we consider the 2D quantization scheme further split into the odd and even cases, see Sec.~\ref{sec:Hamiltonians}) differ from each other distinctly.
The question concerning the validity of Bohigas' conjecture in all quantizations has been raised above.
The answer is evident from the present results:
The Brody parameter for all quantization schemes exhibits essentially the same dependence on energy, the mutual deviations being fully within the range of standard errors.
Therefore, we can report that full agreement with Bohigas' conjecture is confirmed in the present model independently of the chosen quantization scheme.

\section{Conclusions}\label{sec:Conclusions}

We have studied the correspondence between classical and quantum measures of chaos in the geometric collective model adopted from nuclear physics.
In spite of its conceptual simplicity, the model exhibits enormous complexity of solutions, with a very sensitive dependence of the observed behaviors on external parameters and energy.
These features qualify the model for being a subject of detailed analyses of the competition between regular and chaotic modes of motions on both classical and quantum levels.

Although the dependence of chaotic measures on some external control parameters has been extensively studied in quantum billiards, see e.g. Refs.~\cite{Cso93,Li02}, the dependence on energy has been marginalized so far.
This is partly because billiard systems do not permit this kind of analysis---their chaotic features are always uniform in energy.
Even the studies based on \lq\lq soft\rq\rq\ potential systems have been so far focused mostly on the cases with a relatively simple energy dependence of chaotic measures, see e.g. Refs.~\cite{Hen64,Hal84,Sel84}.
The collective Hamiltonians used in Refs.~\cite{Alh91,Whe93,Mac07} and also those studied in Refs.~\cite{Cej04,Str06} provide a very different perspective.
In this sense, the present work can be considered as complementary to the studies based on two-dimensional billiard systems.
We have demonstrated that a tight connection between classical and quantum measures of chaos, embodied in the well-known Bohigas' conjecture \cite{Boh84}, remains valid even under the condition of a strong energy dependence.

Another important conclusion of our work is the observed independence of quantum chaotic measures on the method used to quantize the system.
Since the definition of quantum chaos is based on the system's semiclassical limit, it would be very surprising to find the opposite---i.e. statistical properties of spectra depending on the quantization.
However, the question deserves to be tested.
The present work is probably only a first step in this direction.

Finally, our results show that the Brody parameter, despite of being often deprecated, represents a reasonably sensitive measure of chaos in quantum system.
We nevertheless do not stop at this stage.
In the forthcoming part of this paper \cite{II}, features of the geometric model are analyzed with the aid of the method proposed by Peres \cite{Per84}.
This method enabled us to study the breakdown and reoccurrence of ordered quantal spectra with running parameter $B$ in a much more detailed way.

\acknowledgments
The authors would like to thank M.~Kladiva, M.~Macek, and D.~Bucurescu for fruitful discussions.
This work was supported by the Czech Science Foundation (grant no. 202/06/0363) and by the Czech Ministry of Education (contracts no. 0021620859 and LA 314).

\begin{table*}[tbp]
\caption{Matrix elements of the radial potential terms in the 2D oscillator basis. Matrix elements for other combinations of the oscillator states are zero.}
	\begin{tabular}[b]{r c l}
	\hline\hline
			$\matr{nm}{\beta^{2}}{nm}$      &$=$& $k^{-1}\rb{2n+3m+1}\label{matr2D}$\\
			$\matr{n+1,m}{\beta^{2}}{nm}$   &$=$& $-k^{-1}\sqrt{\rb{n+1}\rb{n+3m+1}}$\\
			$\matr{n,m+1}{\beta^{3}}{nm}$   &$=$& $k^{-3/2}\sqrt{\rb{n+3m+3}\rb{n+3m+2}\rb{n+3m+1}}$\\
			$\matr{n-1,m+1}{\beta^{3}}{nm}$ &$=$& $-3k^{-3/2}\sqrt{n\rb{n+3m+2}\rb{n+3m+1}}$\\
			$\matr{n-2,m+1}{\beta^{3}}{nm}$ &$=$& $3k^{-3/2}\sqrt{n\rb{n-1}\rb{n+3m+1}}$\\
			$\matr{n-3,m+1}{\beta^{3}}{nm}$ &$=$& $-k^{-3/2}\sqrt{n\rb{n-1}\rb{n-2}}$\\
			$\matr{nm}{\beta^{4}}{nm}$      &$=$& $k^{-2}\sb{n\rb{n-1}+\rb{n+3m+1}\rb{5n+3m+2}}$\\
			$\matr{n+1,m}{\beta^{4}}{nm}$   &$=$& $-2k^{-2}\rb{2n+3m+2}\sqrt{\rb{n+3m+1}\rb{n+1}}$\\
			$\matr{n+2,m}{\beta^{4}}{nm}$   &$=$& $k^{-2}\sqrt{\rb{n+3m+2}\rb{n+3m+1}\rb{n+2}\rb{n+1}}$\\
	\hline\hline
	\end{tabular}
\label{tab:Matrix}
\end{table*}

\appendix
\section{Hamiltonian matrix elements}\label{sec:MatrixElements}

Here we expose the 2D and 5D oscillator wave functions used for the diagonalization of both versions of the GCM Hamiltonian for $J=0$ and write down explicit expressions for relevant matrix elements.
There exists a tight connection between the 2D and 5D bases, which shows up particularly in the form of matrix elements.
Note that our derivation of matrix elements is based on Ref.~\cite{Bel70}, but an alternative algebraic approach was recently presented also in Ref.~\cite{Disertace}.
	
We employ the 2D and 5D oscillator bases in radial coordinates $\beta$ and $\gamma$.
The oscillator Hamiltonian is given by $H^{\bullet}_{\rm osc}=T^{\bullet}+A_{\rm osc}\beta^2$, where $\bullet=$2D or 5D and $T^{\bullet}$ is the kinetic energy \eqref{eq:T2D} or \eqref{eq:T5D}.
The oscillator eigenstates in the 2D and 5D cases are denoted as $\ket{nm}$ and $\ket{\nu\mu}$, respectively, where $n$ and $\nu$ represent the corresponding radial quantum numbers.
These states are also eigenstates of the angular momentum associated with the rotations varying angle $\gamma$, therefore they carry good quantum number $m$ corresponding to the O(2) invariant in the 2D case, or $\mu$ corresponding to the O(5) invariant with $J=0$ in the 5D case \cite{Cha76}.
Matrix elements of a general GCM Hamiltonian in these bases can be calculated analytically.
	
Starting with the 2D case, the basis wave functions read as
\begin{equation}
\scal{\beta,\gamma}{nm}\equiv\Psi^{\mathrm{2D}}_{nm}(\beta,\gamma)=R^{\mathrm{2D}}_{nm}(\beta)\Phi^{\mathrm{2D}}_{m}(\gamma)
\end{equation}
with the radial and angular parts given by
\begin{align}
R^{\mathrm{2D}}_{nm}(\beta)
	&=\sqrt{\frac{2kn!}{\rb{n+3m}!}}\rb{k\beta^{2}}^{\frac{3m}{2}}\e^{\frac{-k\beta^{2}}{2}}L^{3m}_{n}(k\beta^{2})
	\nonumber\\
	\Phi^{\mathrm{2D}}_{{\mathrm O}m}(\gamma)
	&=\frac{1}{\sqrt{\pi}}\sin{3m\gamma} \qquad\qquad m=0,1,\dots\nonumber\\
	\Phi^{\mathrm{2D}}_{{\mathrm E}m}(\gamma)
	&=\left\{
	\begin{array}{ll}\frac{1}{\sqrt{2\pi}} & \qquad \textrm{if $m=0$}\\ 
	\frac{1}{\sqrt{\pi}}\cos{3m\gamma}     & \qquad \textrm{if $m=1,2,\dots$}
	\end{array} \right.
\end{align}
Here, $L_n^m$ denotes the Laguerre polynomial and $k=\sqrt{2A_{\rm osc}K}/\hbar$.
The angular part is written for both odd (O) and even (E) cases.
Note that we have applied the symmetry condition \eqref{eq:WFSymmetry1}, selecting only the states with angular momentum quantum numbers equal to multiples of 3.
The states are normalized as follows,
\begin{align}
	&\int_{0}^{\infty}R^{\mathrm{2D}}_{n'm}(\beta)R^{\mathrm{2D}}_{nm}(\beta)\beta\,d\beta\nonumber\\
	&\quad\times
	\int_{0}^{2\pi}\Phi^{\mathrm{2D}}_{{\rm X}m'}(\gamma)\Phi^{\mathrm{2D}}_{{\rm X}m}(\gamma)\,d\gamma
	=\delta_{m'm}\delta_{n'n}
	\,,
\end{align}
with X standing for O or E.

Following Ref.~\cite{Bel70} (where however a different normalization of Laguerre polynomials is used), we can calculate the relevant matrix elements.
First, the oscillator Hamiltonian itself trivially yields
\begin{equation}
\matr{n',m'}{H^{\rm 2D}_{\rm osc}}{nm}=\hbar\Omega(2n+3m+1)\delta_{n'n}\delta_{m'm}
\,,
\label{eq:triv}
\end{equation}
where $\Omega=\sqrt{2A_{\rm osc}/K}$.
To calculate matrix elements of a general Hamiltonian $H^{\rm 2D}$, we need to know matrix elements of the individual terms in $V'=(A-A_{\rm osc})\beta^2+B\beta^{3}\cos{3\gamma}+C\beta^4$.
The radial parts of these elements can be read off from Tab.~\ref{tab:Matrix} and the angular contribution is given by
\begin{equation}
	\matr{n',m+1}{\cos{3\gamma}}{nm}=\left\{
	\begin{array}{ll}\frac{1}{\sqrt{2}} & \quad \textrm{if $m=0$}\\                                                      
			                 \frac{1}{2}    & \quad \textrm{if $m=1,2,\dots$}
	\end{array} \right.
\end{equation}
(for other combinations of basis states, the matrix elements vanish).
	
Let us turn now to the 5D case.
Following the same procedure as above, we express the wave function
\begin{equation}
	\scal{\beta,\gamma}{\nu\mu}\equiv\Psi^{\mathrm{5D}}_{\nu\mu}(\beta,\gamma)
			=R^{\mathrm{5D}}_{\nu\mu}(\beta)\Phi^{\mathrm{5D}}_{\mu}(\gamma)
\end{equation}
with radial and angular components
\begin{align}
	R^{\mathrm{5D}}_{\nu\mu}(\beta)
	&=\sqrt{\frac{2\nu!}{\Gamma\rb{\nu+3\mu+\frac{5}{2}}}}k^{\frac{5}{4}}\rb{k\beta^{2}}^{\frac{3\mu}{2}}
	\e^{\frac{-k\beta^{2}}{2}}L^{3\mu+\frac{3}{2}}_{\nu}(k\beta^{2})
	\nonumber\\
	\Phi^{\mathrm{5D}}_{\mu}(\gamma)
	&=\sqrt{\frac{2\mu+1}{4}}P_{\mu}(\cos{3\gamma})
\end{align}
satisfying the normalization
\begin{align}
		&\int_{0}^{\infty}R^{\mathrm{5D}}_{\nu'\mu}(\beta)R^{\mathrm{5D}}_{\nu\mu}(\beta)\beta^{4}\,d\beta\nonumber\\
		&\quad\times\int_{0}^{2\pi}\Phi^{\mathrm{5D}}_{\mu'}(\gamma)\Phi^{\mathrm{5D}}_{\mu}(\gamma)\abs{\sin{3\gamma}}\,d\gamma
			=\delta_{\mu'\mu}\delta_{\nu'\nu}
\,.
\label{eq:5DNormalization}
\end{align}
In analogy with \eqref{eq:triv} we have
\begin{equation}
\matr{\nu',\mu'}{H^{\rm 5D}_{\rm osc}}{\nu\mu}=\hbar\Omega\left(2\nu+3\mu+\tfrac{5}{2}\right)\delta_{\nu'\nu}\delta_{\mu'\mu}
\,.
\end{equation}
The radial matrix elements can be simply obtained from Tab.~\ref{tab:Matrix} after substitution $n\rightarrow\nu$, 
$m\rightarrow\mu$ on the left-hand side, and $n\rightarrow\nu$, $m\rightarrow\mu+1/2$ on the right-hand side.
The angular part reads as
\begin{equation}
		\matr{\nu',\mu+1}{\cos{3\gamma}}{\nu\mu}
			=\frac{\mu+1}{\sqrt{\rb{2\mu+1}\rb{2\mu+3}}}
\,.
\end{equation}


\end{document}